\documentclass[12pt]{article}

\usepackage[a4paper,margin=1in]{geometry}
\usepackage[T1]{fontenc}
\usepackage[utf8]{inputenc}
\usepackage{newtxtext,newtxmath}
\usepackage{graphicx}
\usepackage{amsmath}
\usepackage{booktabs}
\usepackage{caption}
\usepackage{setspace}
\usepackage[numbers,super,sort&compress]{natbib}
\usepackage[hidelinks]{hyperref}

\graphicspath{{./}{figures/}}
\newenvironment{affiliations}{%
	\begin{list}{}{\setlength{\leftmargin}{0pt}\setlength{\itemindent}{0pt}\setlength{\labelwidth}{0pt}\setlength{\labelsep}{0pt}\setlength{\itemsep}{2pt}\small}
	}{\end{list}}

\title{Lagged sea-surface-temperature precursors of the leading PM$_{2.5}$ mode in China}

\author{Yuan Chen$^{1,2\dagger}$, Dan Zhao$^{1,2\dagger}$, Xu Li$^{1,2\ast}$}

\date{}

\begin{document}
	\maketitle
	
	\begin{affiliations}
		\item $^{1}$Yunnan Key Laboratory of Complex Systems and Brain-Inspired Intelligence, Kunming University of Science and Technology, Kunming 650500, Yunnan, China.
		\item $^{2}$Faculty of Science, Kunming University of Science and Technology, Kunming 650500, Yunnan, China.
		\item $^{\dagger}$These authors contributed equally: Yuan Chen and Dan Zhao.
		
		\item $^{\ast}$Correspondence and requests for materials should be addressed to Xu Li.\\
		E-mail: \texttt{lixucn@kust.edu.cn}.

	\end{affiliations}
	
	\vspace{2em}

	\noindent\textbf{
		Fine particulate matter (PM$_{2.5}$) pollution in China is strongly modulated by meteorological variability, yet its seasonal predictability from oceanic signals remains unclear. Here we identify the leading PM$_{2.5}$ variability mode over China and show that it is preceded by coherent sea-surface-temperature anomaly clusters by more than one season. These oceanic precursors influence summer PM$_{2.5}$ mainly by altering precipitation and low-level ventilation, and winter PM$_{2.5}$ by modulating boundary-layer height and near-surface stagnation. Using the four largest precursor regions, a simple regression model achieves significant independent prediction skill for both summer and winter PM$_{2.5}$ variability. Our results reveal a physical pathway linking sea-surface-temperature memory to regional aerosol pollution and provide a basis for seasonal air-quality risk assessment.
	}
	
	\vspace{2em}
	
	\section*{Introduction}
	
	Fine particulate matter (PM$_{2.5}$) is a leading environmental risk factor for premature mortality because it penetrates deeply into the respiratory system and increases cardiopulmonary and respiratory disease risks\cite{Pope2002,Cohen2017Lancet,Burnett2018}. In China, severe PM$_{2.5}$ pollution has historically occurred over densely populated and industrialized regions, including the North China Plain, the Yangtze River Delta and the Sichuan Basin. Clean-air policies have substantially reduced PM$_{2.5}$ concentrations since 2013\cite{Zhang2019PNAS}, but large interannual variability and episodic pollution potential remain strongly modulated by meteorological conditions and climate variability\cite{Zhai2019ACP,Yu2023}.
	
	Pollution formation and persistence depend on the balance between emissions, chemical production, deposition and atmospheric dispersion. Local meteorological factors such as wind speed, humidity, precipitation and planetary boundary-layer height regulate the accumulation and removal of aerosols\cite{Seinfeld2016,Petaja2016SciRep,Zhong2018}. A shallow boundary layer suppresses vertical dilution, weak winds reduce horizontal ventilation, and precipitation removes particles through wet scavenging. These local processes are well established, but they do not fully explain why PM$_{2.5}$ anomalies can vary coherently over broad regions of China. Such spatial coherence suggests a role for large-scale circulation anomalies that simultaneously alter ventilation and removal processes across multiple polluted regions.
	
	Sea surface temperature (SST) is a key source of low-frequency climate memory. Owing to the large heat capacity of the ocean mixed layer, SST anomalies can persist for months and act as slowly varying boundary forcing for the atmosphere\cite{Hasselmann1976,Monetti2003PhysA_SST}. Through air--sea heat and moisture exchanges, SSTA can induce anomalous diabatic heating, excite planetary-scale circulation responses and modulate Rossby-wave-like teleconnections\cite{RossbyWang2013,Zhang2019grl}. These processes can influence distant atmospheric conditions over East Asia, including monsoon rainfall, cold-air activity and boundary-layer stability. Previous work has demonstrated that ocean-memory signals can improve seasonal prediction of aerosol pollution in South Asia\cite{Gao2019SciAdv_OceanMemory}. For China, however, the direct connection between SSTA precursor regions, national-scale PM$_{2.5}$ modes and the meteorological pathways of influence requires further clarification.
	
	Here we analyse PM$_{2.5}$ pollution in China as a high-dimensional complex system whose macroscopic variability can be represented by a small number of dominant modes\cite{Fan2021PR,Hu2019SCPMA}. Rather than relying on predefined climate indices, we directly identify SSTA regions that significantly precede the leading PM$_{2.5}$ mode and then examine the associated precipitation, wind and boundary-layer responses. This design focuses on the physical pathway by which oceanic thermal anomalies influence PM$_{2.5}$: SSTA modifies large-scale atmospheric circulation, circulation reshapes regional meteorology, and the resulting meteorological conditions regulate particle removal, mixing and accumulation.
	
	\section*{Results}
	
	\subsection*{Dominant spatiotemporal mode of PM$_{2.5}$ pollution}
	
\begin{figure}[htbp]
		\centering
		\includegraphics[width=0.98\linewidth]{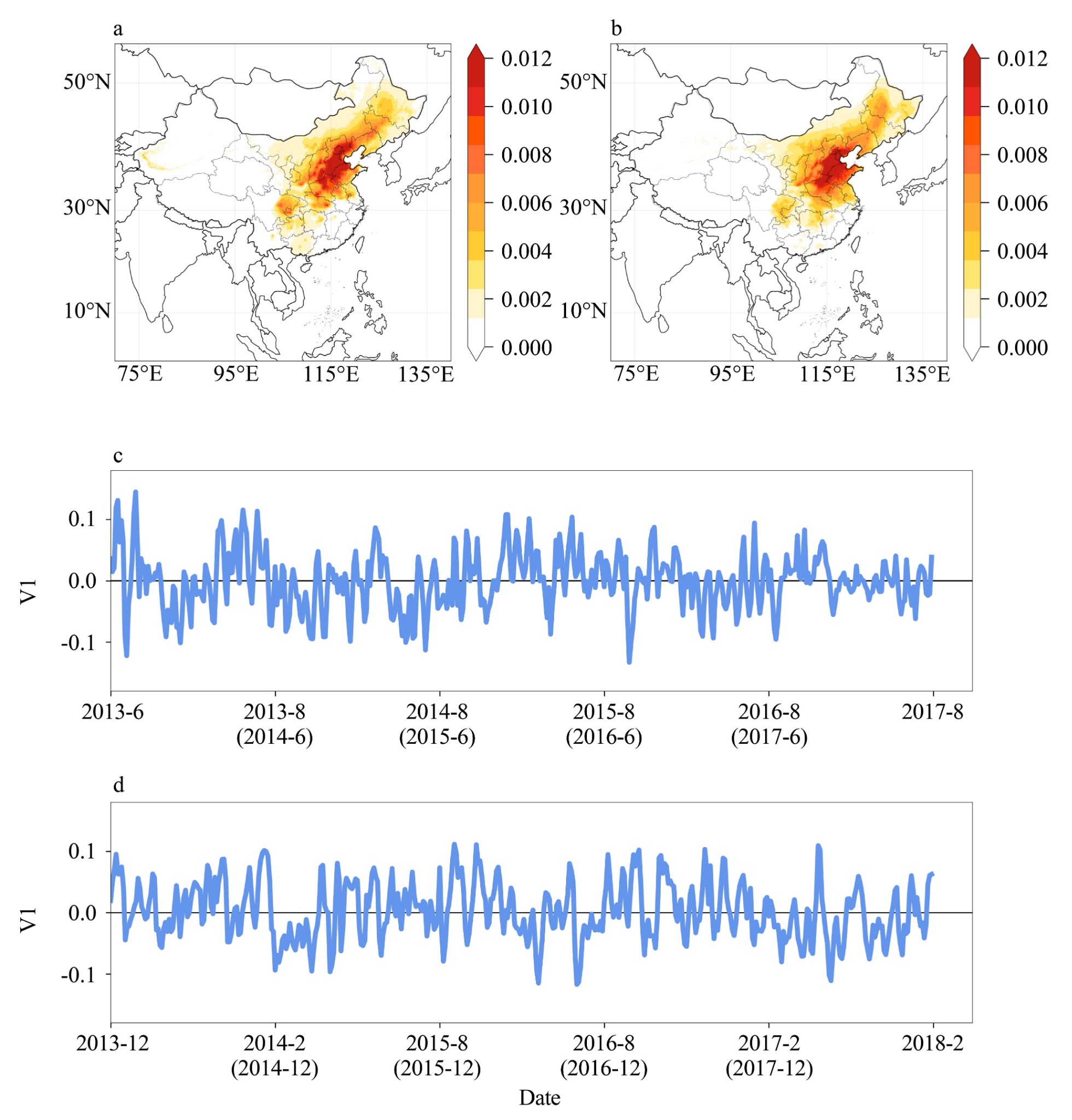}
		\caption{\textbf{Dominant PM$_{2.5}$ eigen-microstate in China.} Spatial distribution of the first eigen-microstate in summer and winter, together with its temporal evolution coefficient $V_1$. The first mode describes the coherent PM$_{2.5}$ anomaly pattern over eastern China and explains 18.1\% of summer variance and 31.4\% of winter variance.}
		\label{fig:em1}
	\end{figure}

	The leading eigen-microstate of PM$_{2.5}$ captures the primary coherent pattern of pollution variability in China (Fig.~\ref{fig:em1}). In both summer and winter, the strongest loadings occur over eastern China, especially the North China Plain and surrounding densely populated regions. This structure indicates that the dominant mode is not a purely local signal, but a broad-scale pollution pattern controlled by shared meteorological background conditions.
	
	The explained variance differs strongly by season. The first eigen-microstate explains 18.1\% of total PM$_{2.5}$ variance in summer and 31.4\% in winter. The larger winter contribution suggests that cold-season pollution is more strongly organized by large-scale atmospheric circulation. During winter, weak synoptic ventilation, reduced turbulent mixing, frequent stable stratification and shallow boundary layers can affect wide regions simultaneously. In summer, stronger convection, more frequent precipitation and enhanced local chemical and boundary-layer variability reduce the spatial coherence of PM$_{2.5}$ anomalies, leading to a lower explained variance.
	
	The temporal coefficient $V_1$ represents the daily evolution of the dominant PM$_{2.5}$ mode. Positive phases correspond to enhanced PM$_{2.5}$ anomalies over the main polluted regions, whereas negative phases correspond to cleaner conditions. Because $V_1$ condenses the national-scale PM$_{2.5}$ field into a physically interpretable collective variable, it is used as the target for the following SSTA teleconnection and prediction analyses.

	\subsection*{Lagged SSTA precursor regions}
	
\begin{figure}[htbp]
		\centering
		\includegraphics[width=0.98\linewidth]{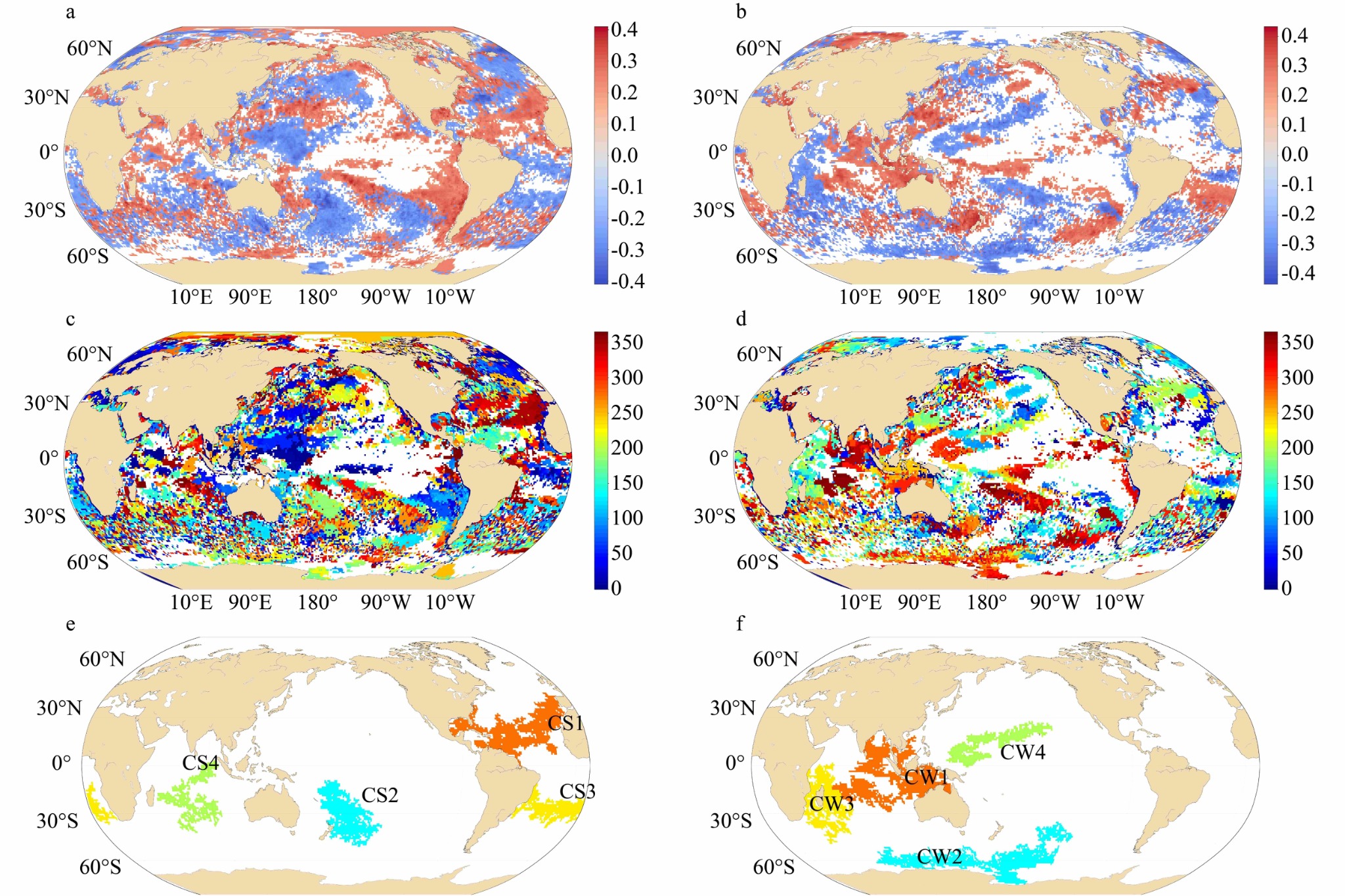}
		\caption{\textbf{Lagged SSTA regions associated with the dominant PM$_{2.5}$ mode.} Spatial distributions of significant lagged correlation coefficients between global SSTA and $V_1$, corresponding lag times, and the four largest SSTA precursor clusters in summer and winter. White areas denote grid points that do not pass the significance test. The selected clusters lead the PM$_{2.5}$ mode by more than 90 days.}
		\label{fig:ssta}
	\end{figure}

	Global SSTA fields show significant lagged correlations with the leading PM$_{2.5}$ coefficient $V_1$ (Fig.~\ref{fig:ssta}). The significant regions are not randomly distributed; instead, they form coherent oceanic clusters. Many of these clusters precede $V_1$ by more than 90 days, indicating that they contain seasonal-scale precursor information rather than merely reflecting simultaneous atmospheric noise.
	
	The physical relevance of these SSTA clusters arises from ocean thermal inertia. Persistent warm or cold anomalies in the ocean mixed layer alter the surface energy balance and moisture fluxes over key ocean basins. These flux anomalies can modify atmospheric heating patterns, shift large-scale circulation and create remote responses over East Asia. The lagged relationship therefore suggests that the ocean acts as an integrator of atmospheric variability and provides a slow boundary condition that affects later PM$_{2.5}$ dispersion environments.
	
	For prediction, the four largest significant SSTA clusters in each season were selected after requiring statistical significance and a lag time of at least 90 days. These clusters were used directly as oceanic precursor regions. No climate-index matching was applied, because the goal is to retain the spatial information of the oceanic forcing itself and to connect it directly to downstream meteorological fields over China.

	\subsection*{Summer pathway: precipitation and horizontal ventilation}
	
\begin{figure}[htbp]
		\centering
		\includegraphics[width=0.98\linewidth]{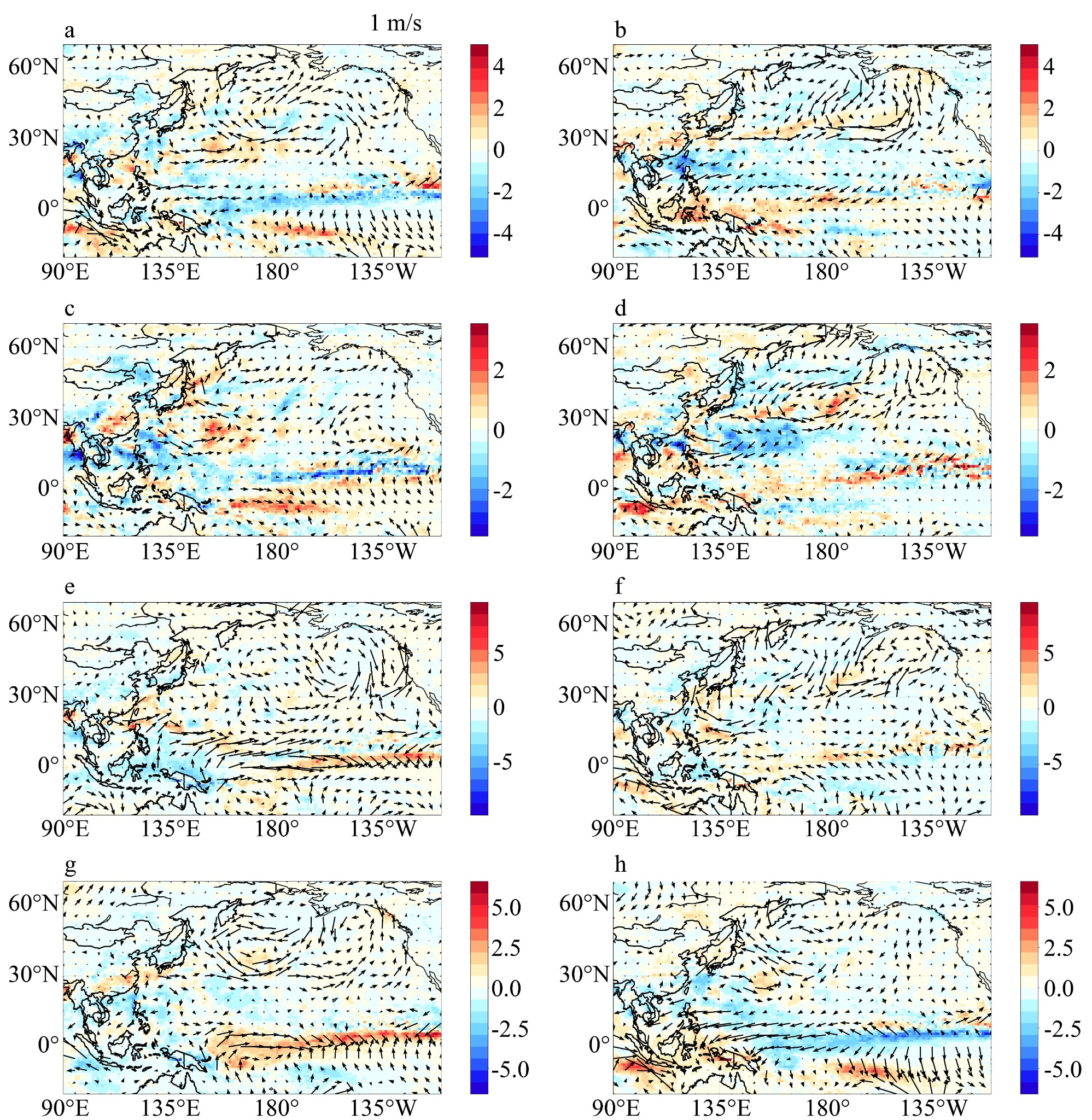}
		\caption{\textbf{Summer meteorological pathway linking SSTA to PM$_{2.5}$.} Lagged composite precipitation and 10-m wind anomalies associated with SSTA precursor states. Shading denotes precipitation anomalies, and vectors denote 10-m wind anomalies. Pollution-favourable states are characterized by reduced wet removal and/or weakened horizontal ventilation over key polluted regions.}
		\label{fig:summer}
	\end{figure}

	The summer composites reveal how oceanic thermal anomalies are translated into PM$_{2.5}$-relevant meteorological conditions (Fig.~\ref{fig:summer}). Under SSTA precursor conditions associated with enhanced summer PM$_{2.5}$, precipitation anomalies over eastern China display a spatial redistribution, with reduced rainfall over major northern pollution regions in several composite states. At the same time, anomalous low-level wind fields weaken horizontal ventilation or produce circulation patterns that favour pollutant retention.
	
	This pathway is physically direct. Summer PM$_{2.5}$ is strongly regulated by wet deposition because monsoonal rainfall efficiently scavenges particles from the lower troposphere. When SSTA-related circulation anomalies shift moisture transport and suppress precipitation over northern and eastern China, the wet removal efficiency decreases. Weak or unfavourable 10-m wind anomalies further reduce horizontal dispersion. These two effects jointly increase the persistence of PM$_{2.5}$ anomalies in the dominant pollution regions.
	
	The opposite SSTA-related composite states are associated with enhanced precipitation or more favourable ventilation over polluted regions, corresponding to reduced PM$_{2.5}$ anomalies. Thus, in summer, the dominant mechanism is not direct oceanic forcing of aerosols, but a chain in which oceanic thermal anomalies reshape monsoon-season rainfall and near-surface transport, and those meteorological changes regulate particle removal and dispersion.

	\subsection*{Winter pathway: boundary-layer suppression and stagnation}
	
\begin{figure}[htbp]
		\centering
		\includegraphics[width=0.98\linewidth]{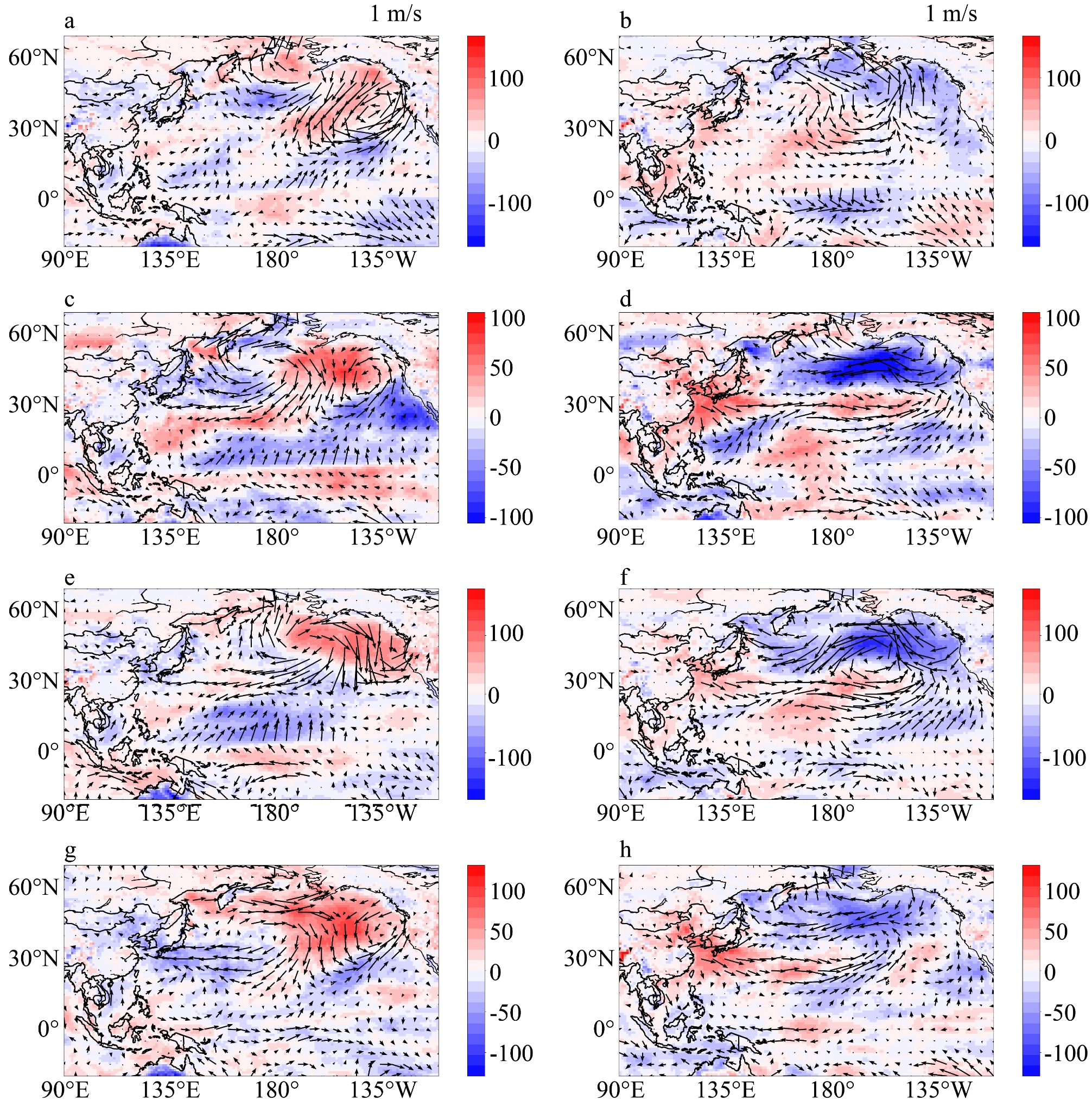}
		\caption{\textbf{Winter meteorological pathway linking SSTA to PM$_{2.5}$.} Lagged composite boundary-layer-height and 10-m wind anomalies associated with winter SSTA precursor states. Shading denotes boundary-layer-height anomalies, and vectors denote 10-m wind anomalies. Pollution-favourable states are characterized by lower boundary-layer height and weak near-surface ventilation over eastern China.}
		\label{fig:winter}
	\end{figure}

	The winter composites indicate a different mechanism (Fig.~\ref{fig:winter}). SSTA precursor states associated with enhanced winter PM$_{2.5}$ are followed by negative boundary-layer-height anomalies over eastern China, especially from the North China Plain to the Yangtze River Delta. Near-surface wind anomalies are weak or arranged in a way that limits effective ventilation. Together, these anomalies create stagnant lower-tropospheric conditions.
	
	The winter pathway is dominated by vertical mixing and atmospheric stability. A lower boundary layer compresses the vertical volume available for pollutant dilution. When aerosols and their precursors are confined within a shallower layer, near-surface PM$_{2.5}$ concentrations increase. Weak surface winds simultaneously suppress horizontal transport, while stable stratification favours the persistence of accumulated particles. SSTA-related circulation anomalies can therefore amplify the background winter tendency toward stagnation by strengthening subsidence, suppressing turbulent mixing and weakening cold-air ventilation.
	
	This explains why the leading PM$_{2.5}$ mode is more coherent in winter than in summer. When the large-scale circulation favours shallow boundary layers and weak winds, many polluted regions respond in phase. The ocean-memory signal is thus expressed through boundary-layer dynamics and stagnation rather than through wet deposition.

	\subsection*{Seasonal prediction using SSTA clusters}
	
\begin{figure}[htbp]
		\centering
		\includegraphics[width=0.98\linewidth]{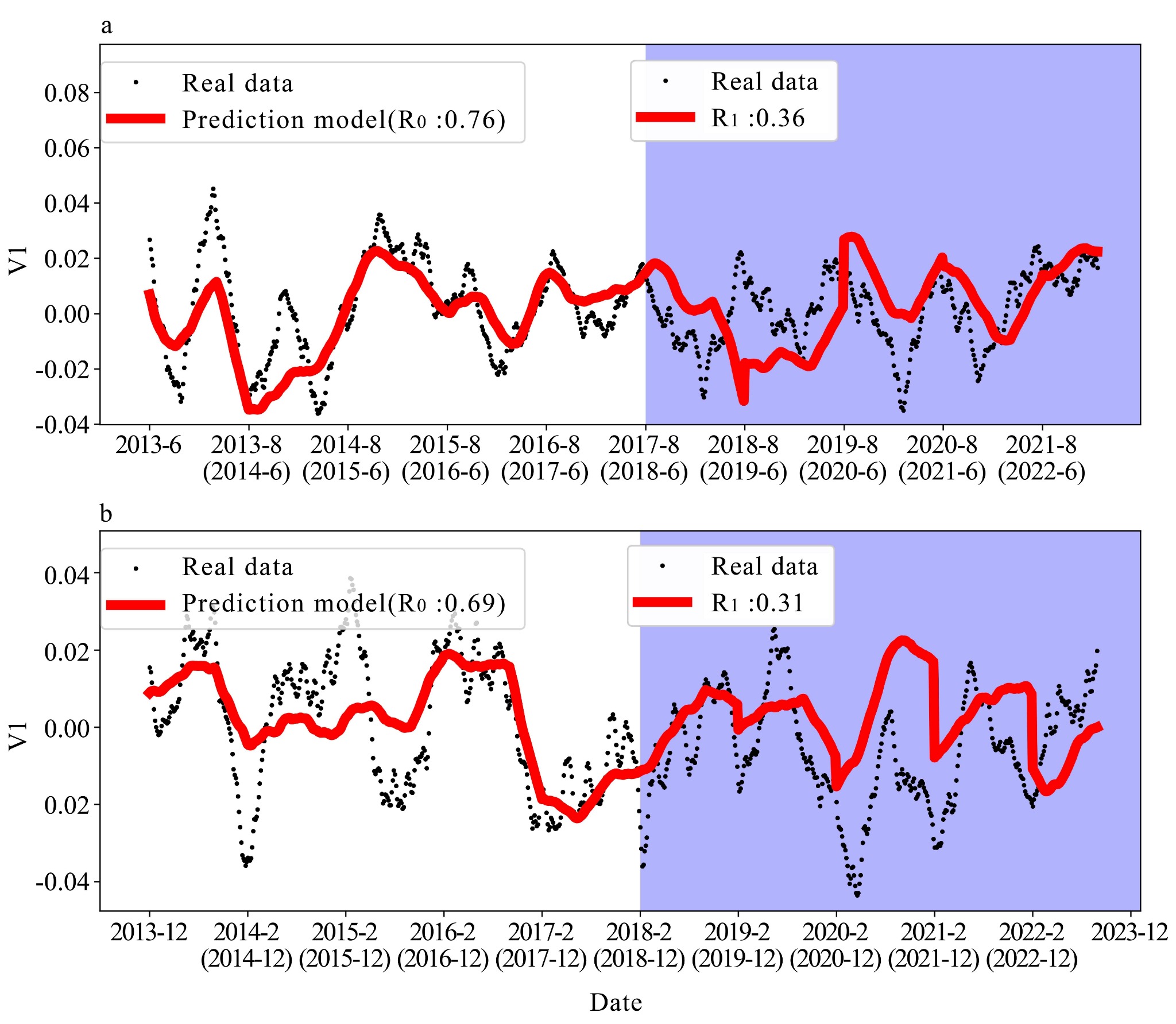}
		\caption{\textbf{Seasonal prediction of the dominant PM$_{2.5}$ mode.} Observed and predicted 30-day running-mean $V_1$ in summer and winter. The regression model uses the four largest lagged SSTA clusters as predictors. Prediction skill remains significant in the independent testing period, with correlations of 0.36 in summer and 0.31 in winter.}
		\label{fig:prediction}
	\end{figure}

	The four largest SSTA precursor clusters were used to predict the 30-day running mean of $V_1$ through multivariate linear regression (Fig.~\ref{fig:prediction}). In the training period, the predicted and observed $V_1$ values are strongly correlated, with $R=0.76$ in summer and $R=0.69$ in winter. In the independent testing period, prediction skill decreases, as expected for seasonal air-quality prediction, but remains statistically significant, with $R=0.36$ in summer and $R=0.31$ in winter.
	
	The reduced skill in the independent period reflects the complexity of PM$_{2.5}$ variability. Emission changes, local weather noise, aerosol chemistry and aerosol--boundary-layer feedbacks can all weaken the relationship between remote SSTA and local pollution. Nevertheless, the significant testing skill demonstrates that SSTA contains useful information about the low-frequency background state of PM$_{2.5}$ pollution. The model should therefore be interpreted as a seasonal pollution-potential predictor rather than a short-term event forecast.
	
	Because the selected SSTA clusters lead $V_1$ by more than 90 days, this framework extends prediction beyond the time scale of conventional weather forecasts. Such lead time is useful for seasonal air-quality risk assessment and for designing proactive regional pollution-control strategies before seasons with unfavourable dispersion potential.

	\section*{Discussion}
	
	Our results show that SSTA provides physically interpretable precursor information for the dominant PM$_{2.5}$ pollution mode in China. The influence of SSTA is indirect but systematic: persistent oceanic thermal anomalies alter large-scale atmospheric circulation, the circulation response changes regional meteorological dispersion conditions, and these conditions determine whether particles are removed, diluted or accumulated.
	
	The seasonal contrast is central to this mechanism. In summer, SSTA affects PM$_{2.5}$ mainly by modulating precipitation and low-level winds. Reduced rainfall weakens wet deposition, and unfavourable wind anomalies limit horizontal ventilation. In winter, SSTA affects PM$_{2.5}$ mainly by changing boundary-layer height and surface ventilation. A shallow and stable boundary layer confines pollutants near the surface, and weak winds prevent dispersion. These pathways are consistent with established knowledge of aerosol removal, boundary-layer control and haze persistence\cite{Seinfeld2016,Zhong2018,Su2020,Ma2020}.
	
	The direct SSTA-cluster approach has two advantages. First, it avoids dependence on predefined climate indices and therefore preserves the geographic structure of oceanic thermal forcing. Second, it provides a transparent link from precursor regions to meteorological fields, allowing the mechanism to be interpreted through precipitation, wind and boundary-layer responses. This is particularly important for air-quality prediction, where a useful climate precursor must be connected to the actual meteorological processes controlling pollutant concentration.
	
	Several limitations remain. The PM$_{2.5}$ record beginning in 2013 is short relative to the time scale of climate variability, and longer records will be needed to test the stability of the SSTA--PM$_{2.5}$ relationship. The regression model is deliberately simple and interpretable, but it cannot represent nonlinear aerosol chemistry, emission-policy changes or high-frequency weather variability. In addition, the target variable is a national-scale dominant mode rather than local station-level PM$_{2.5}$. Future work should combine SSTA-based climate predictors with dynamical forecasts, emission inventories and nonlinear statistical models to improve regional and event-scale prediction.
	
	\section*{Methods}
	
	\subsection*{Datasets}
	
	Daily near-surface PM$_{2.5}$ concentration data over China were analysed from June 2013 to February 2023. Global SST was obtained from the ERA5 reanalysis\cite{Hersbach2020}. Four daily analysis times were averaged to form daily SST fields, which were then regridded to 1$^{\circ}\times$1$^{\circ}$ resolution to focus on basin-scale oceanic variability. The study focuses on summer and winter because these seasons represent contrasting PM$_{2.5}$ regimes: summer pollution is strongly influenced by monsoonal precipitation and wet deposition, whereas winter pollution is dominated by boundary-layer stability and stagnation.
	
	\subsection*{Anomaly construction}
	
	For both PM$_{2.5}$ and SST, daily anomaly series were constructed at each grid point. First, a linear trend was removed:
	\begin{equation}
		X_i^{dt}(t)=X_i(t)-\left(\alpha_i+\beta_i t\right),
	\end{equation}
	where $X_i(t)$ is the original value at grid point $i$, and $\alpha_i$ and $\beta_i$ are the fitted intercept and trend. The mean seasonal cycle was then removed by subtracting the multi-year daily climatology:
	\begin{equation}
		X_i'(t)=X_i^{dt}(t)-\frac{1}{Y}\sum_{y=1}^{Y}X_i^{dt}(y,d),
	\end{equation}
	where $d$ is the calendar day and $Y$ is the number of available years. Leap days were removed so that each year contained 365 days. This procedure reduces trend- and seasonality-induced spurious correlations.
	
	\subsection*{Extraction of PM$_{2.5}$ eigen-microstates}
	
	For each season, the PM$_{2.5}$ anomaly field was arranged into a space--time matrix $\mathbf{X}$ with dimension $M\times N$, where $M$ is the number of grid points and $N$ is the number of daily samples. The covariance matrix of the spatial grid points was decomposed into eigenvalues and eigenvectors. The eigenvector with the largest eigenvalue was defined as the first eigen-microstate (EM1), and the projection of the original field onto this eigenvector was defined as its temporal evolution coefficient $V_1(t)$. EM1 represents the leading collective mode of PM$_{2.5}$ variability.
	
	\subsection*{Lagged SSTA correlation and cluster selection}
	
	For each global SST grid point $i$, the normalized lagged correlation between SSTA $S_i(t)$ and the PM$_{2.5}$ coefficient $V_1(t)$ was calculated as
	\begin{equation}
		CC_{V_1 i}(\tau)=\frac{\langle V_1(t)S_i(t-\tau)\rangle}{\sqrt{\langle V_1(t)^2\rangle\langle S_i(t-\tau)^2\rangle}},
	\end{equation}
	where $\tau$ is the lag time from 0 to 365 days. Positive $\tau$ indicates that SSTA leads PM$_{2.5}$. For each grid point, the maximum absolute correlation and its corresponding lag were retained. Statistical significance was assessed using a Student's $t$ test at the 95\% confidence level. Grid points were retained only if they passed the significance test and had a lag time of at least 90 days.
	
	Spatially adjacent retained grid points were grouped into connected SSTA clusters using four-neighbour connectivity. The effective area of each cluster was calculated using latitude weighting:
	\begin{equation}
		S_k=\sum_{g\in\Omega_k}\cos(\phi_g),
	\end{equation}
	where $\Omega_k$ denotes the grid points in cluster $k$ and $\phi_g$ is latitude. The four clusters with the largest effective areas were selected as seasonal oceanic predictors.
	
	\subsection*{Meteorological composites and prediction model}
	
	Meteorological composites were constructed to diagnose how SSTA precursor states influence PM$_{2.5}$-relevant atmospheric conditions. For summer, precipitation and 10-m wind anomalies were analysed to evaluate wet deposition and horizontal ventilation. For winter, boundary-layer height and 10-m wind anomalies were analysed to evaluate vertical dilution and stagnation. Composite states were defined from the SSTA precursor clusters and compared between pollution-favourable and pollution-unfavourable conditions.
	
	For prediction, each selected SSTA cluster was converted into a daily predictor by calculating the correlation-weighted mean SSTA over the cluster. The predictors were shifted according to their optimal lag times and smoothed with a 30-day running mean. A multivariate linear regression model was then fitted:
	\begin{equation}
		\overline{V_1}(t)=\sum_{k=1}^{4}a_k\overline{SST'_k}(t)+b,
	\end{equation}
	where the overbar denotes 30-day smoothing, $SST'_k(t)$ is the lag-adjusted SSTA predictor for cluster $k$, $a_k$ are regression coefficients, and $b$ is the intercept. The first five years were used for training, and the following years were used for independent testing. Prediction skill was evaluated using Pearson correlation, with significance assessed by a Student's $t$ test.
	
	\section*{Data availability}
	The ERA5 reanalysis data are publicly available from the Copernicus Climate Data Store. The PM$_{2.5}$ dataset used in this study is available from its data provider. Processed data supporting the findings of this study can be made available from the corresponding author upon reasonable request.
	
	\section*{Code availability}
	The analysis code used for anomaly construction, eigen-microstate extraction, lagged correlation analysis and regression prediction can be made available from the corresponding author upon reasonable request.
	
	\section*{Acknowledgements}
	The authors acknowledge the providers of the PM$_{2.5}$ dataset and ERA5 reanalysis. This work benefited from discussions on complex systems, climate teleconnections and air-quality prediction.
	
	\section*{Author contributions}
	Y.C. designed the study, performed the analysis and drafted the manuscript. D.Z. contributed to data processing, methodological development and interpretation of the results. X.L. supervised the study, contributed to the physical interpretation and revised the manuscript. All authors reviewed the manuscript.

\end{document}